\documentclass[twocolumn]{revtex4}
\usepackage{graphicx,amsmath,amssymb,mathrsfs}
\usepackage{epsfig}

\begin{document}

\title{Possible resonance effect of axionic dark matter in Josephson junctions}

\author{Christian Beck}

\affiliation{Isaac Newton Institute for Mathematical Sciences, University of Cambridge,
20 Clarkson Road, Cambridge CB3 0EH, UK, and \\
School of Mathematical Sciences, Queen Mary University of London, Mile End Road, London E1 4NS, UK}

\begin{abstract}
%We discuss a novel interaction process of dark matter axions in resonant
%S/N/S Josephson junctions, which has similarities with an axionic Josephson
%effect. Axions briefly decay into two microwave
%
%
%We discuss the possibility
We provide theoretical arguments
that dark matter axions from the galactic halo that pass through the earth
may generate a small observable signal in resonant S/N/S Josephson junctions.
%&&&
%by briefly decaying into microwave
%photons
%under the influence of resonant Josephson radiation and
%recombining back
%into an axion.
%&&&&
%
%In total this process triggers one additional Cooper pair to tunnel,
%which produces a measurable additional supercurrent if the Josephson frequency
%$\omega_J$ matches the axion mass $m_ac^2=\hbar \omega_J$.
The corresponding interaction process is based on uniqueness
of the gauge-invariant axion Josephson phase angle modulo $2\pi$
and is predicted to produce
a small Shapiro step-like feature without externally applied microwave radiation
when the Josephson frequency resonates with the axion mass.
%induced by a Primakoff effect that is resonantly enhanced by Josephson radiation.
%Axions
%and a resonantly enhanced Primakoff effect.
%entering the weak-link region of the junction trigger an additional flow of
%Cooper pairs which in principle can be measured.
A resonance signal of so far unknown origin observed in
[C. Hoffmann et al. PRB 70, 180503(R) (2004)] is consistent with our theory and can be interpreted in terms of an
axion mass $m_ac^2=0.11 {\mbox{meV}}$ and a local galactic axionic dark matter density
of $0.05$ $\mbox{GeV}/cm^3$. We discuss future experimental checks to confirm the dark-matter nature of the
observed signal.
\end{abstract}
%We propose new systematic experimental dark matter
%axion searches that look for small yearly modulation effects in measured I-V curves
%of resonant Josephson junctions.
%We suggest a future experiment
%to either confirm or refute the dark matter nature of the observed signal.
%To confirm this dark matter interpretation
%of the observed signal, we suggest to repeat Koch et al.'s experiment and check for a slight yearly periodic
%
%modulation of the resonance signal, similar to that observed in the DAMA/Lyra experiment.

%\begin{abstract}
%We point out that that the equation of motion of the axion is identical to that of a resistively
%shunted Josephson junction (RSJ). This opens up the possibility to build up analogue experiments that
%simulate axionic dark matter in the early and present universe using RSJs. The critical current
%of the RSJ can be tuned to the axion mass $m_a$, the product of the shunt resistance and capacity to
%the Hubble parameter $H$, and the coupling constant $f_a$ of the axion to other fields can be varied
%by varying the strength of the bias current of the RSJ.
%Interestingly, the allowed parameter values of axionic dark matter, as restricted by
%astrophysical and experimental searches,
%can be realized with current Josephson junction technology. We point out several applications of this
%approach.

\maketitle

The existence of dark matter in the universe is one of the major puzzles
of current research in astrophysics, cosmology, and particle physics.
While there is clear evidence for the existence of dark matter
from astronomical
observations, it is still unclear what the physical nature of dark matter is.
Important candidate particles for dark matter are weakly interacting massive particles (WIMPS)
\cite{wimp}
and axions \cite{peccei,duffy}.
Many experimental searches to detect WIMPS \cite{lab1}
and axion-like particles \cite{lab2,lab3,lsw1,lsw2,admx1,graham,jaeckel}
are currently under way or are being discussed for future
implementation. A positive result would be a major breakthrough
in our understanding of the matter contents of the universe.
%Although a few interesting signals have been seen \cite{}
%that could possibly be interpreted in terms of dark matter particles
%(WIMPS in case of \cite{x} and axions in case of \cite{y},
%one is still far away from a generally accepted
%experimental confirmation of
%dark matter in laboratory experiments, since the above signals contradict
%the non-observation of any signal in other experiments.

%Most experiments searching for axions are based on the Primakoff effect,
%i.e. the decay of axions into two microwave photons in the presence of a strong
%magnetic field, as well as the inverse process. For example, in `Light shining through wall' (LSW)
%experiments one looks for photons that convert to axions in a strong magnetic field,
%travel through a `wall', and convert back to photons \cite{lsw1,lsw2}. In other experiments,
%such as the ADMX experiment \cite{admx1}, one looks for axions that decay into microwave
%photons in a resonant cavity.
%But there are also other experimental setups involving axions or axion-like particles
%that can be considered for possible detection processes (see, e.g. \cite{}
%for new ideas in this direction).
%New ideas and novel suggestions how to detect axions, axion-like particles and hidden photons are
%currently developing \cite{graham,jaeckel}.

In this letter we propose a new approach how
to detect
QCD-axionic dark matter in the laboratory with high efficiency,
exploiting a macroscopic quantum effect. Our proposal is based
on S/N/S (Superconductor/Normal Metal/Superconductor) Josephson junctions
as suitable detectors \cite{hoffmann,lhotel,hoss,dubos,tinkham}.
%The effect also occurs in tunnel junctions but it is particularly
%strong in S/N/S junctions, leading to observable signals.
%In short, we may call this effect the `axionic Josephson effect'.
We will provide theoretical arguments
that axions that pass through the weak link region of such a
Josephson junction in the voltage stage may trigger the transport of
additional Cooper pairs if the Josephson frequency $\omega_J$ coincides
with the axion mass $m_ac^2=\hbar \omega_J$.
%For this effect to be most suitable for axion detection one needs junctions with a large
%tunneling area or weak-link region.
The effect is resonantly enhanced.
The basic theoretical idea of our proposal can be regarded as
being a kind of a complement of `Light shining through wall' (LSW) experiments
\cite{lsw1,lsw2}. In LSW photons decay into axions in a strong
magnetic field, which then pass a `wall' and decay back into photons, which can be detected.
Here we employ the opposite effect where axions convert into photons
in a Josephson junction and back into axions when leaving the junction.
%Axions from the galactic halo that enter the weak link region of the S/N/S junction cannot continue to
%exist at the normal metal surface, due to a huge (formal) magnetic field that is simulated
%to them by the driven Josephson junction environment via a phase synchronisation
%effect \cite{beck1, beck2}. They decay into microwave photons and, mediated by multiple
%Andreev reflections of electron hole pairs, tunnel through the metal region to recombine back
%into axions at the other end of the weak link region. This process triggers the transport of
%several additional Cooper pairs through the junction and it creates a measurable signal in
%the differential conductivity of the junction when the axion mass $m_ac^2$ resonates with the
%Josephson frequency $\hbar \omega_J$.
For brevity we may call this effect ATJ
(`Axions tunneling a junction').
The `wall' for axions in this case is represented by the weak-link region of the
biased Josephson junction in the voltage stage.

From an experimental point of view, ATJ predicts a Shapiro-step like feature \cite{shapiro}
in the measured I-V curve of the S/N/S junction that occurs {\em without} externally applied
microwave radiation \cite{he,beck1,beck2}.
The measured differential conductance (see e.g. \cite{hoffmann,lhotel,dubos,hoss} for typical measurement techniques) is predicted to exhibit a small peak at Josephson
frequency $\hbar \omega_J =2eV=m_ac^2$, whose intensity depends on the velocity of
galactic axions hitting the earth, the size of the weak-link region of the junction,
and the local galactic halo density of axions.
%$V$ is the bias voltage. The width of the peak
%is not determined by the velocity dispersion of the axions (they are in
%good approximation monochromatic) but by the line width of the Josephson junction and the
%number of Andreev reflections.

Our calculations in this paper show that the effect of axionic dark matter on the I-V curve is small but observable.
We will discuss a possible candidate signal of
unknown origin that has been observed in \cite{hoffmann},
which interpreted in terms of ATJ provides a prediction of the axion mass of $m_ac^2=0.11$ meV and
an estimate of the local
halo density of dark matter axions of 0.05 GeV/$cm^3$.

%Remarkably,
%a measurement performed by Hoffmann et al. \cite{hoffmann} some 10 years ago (aimed at better understanding
%the noise characteristics of S/N/S junctions) provides experimental evidence for a
%peak consistent with our theoretical prediction. The authors
%of the above paper observed a small temperature-independent
%esonance peak of the differential conductivity at a voltage of 0.055 meV
%for all temperatures below 1K and remark in their paper that `the origin of the additional peak
%at $V\approx 0.06mV$
%is not clear'.
%We here propose to interpret their measured peak in terms of ATJ, i.e. axions hitting the normal metal
%region of their S/N/S junction. From this interpretation
%we obtain a prediction of the axion mass of $m_ac^2=0.11$ meV. From the
%intensity of their observed signal we can also estimate the local
%halo density of dark matter axions as $0.05$ GeV/$cm^3$, which is in agreement with astrophysical
%expectations \cite{dm-density}. Axions with the above mass value are expected to account for about 20$\%$ of the
%dark matter contents of the universe.

%Let us now develop our theory (based on a coherent field effect \cite{graham}) in more detail.
Consider an axion field $a=f_a \theta$, where $\theta$
is the axion misalignment angle and $f_a$ is
the axion coupling constant.
%The equation of motion of the axion misalignment angle $\theta$ in a
%Robertson-Walker metric is
%\begin{equation}
%\ddot{\theta} +3H \dot{\theta}+ \frac{m_a^2c^2}{\hbar^2} \sin \theta = 0. \label{e1}
%\end{equation}
%Here $H$ is the Hubble parameter and $m_a$ denotes the axion mass. The forcing term $\sin \theta$
%is produced by QCD instanton effects. In a mechanical analogue, the
%above equation is that of a pendulum in a constant gravitational field
%with some friction determined by $H$.
%
If strong external electric and magnetic fields
$\vec{E}$ and $\vec{B}$ are present, one has
%then the classical equation
%of motion of the axion misalignement angle is
\begin{equation}
\ddot{\theta} +\Gamma \dot{\theta}+ 
\frac{m_a^2c^4}{\hbar^2} \sin \theta = -\frac{g_\gamma}{4 \pi^2}
\frac{1}{f_a^2} c^3 e^2 \vec{E} \vec{B} \label{1}
\end{equation}
%$g_\gamma$ is a model-dependent dimensionless coupling constant
($g_\gamma =-0.97$ for KSVZ axions \cite{ksvz1,ksvz2},
whereas $g_\gamma=0.36$ for DFSZ axions \cite{dfsz1,dfsz2}). $\Gamma$ is a damping parameter. In the early
universe, $\Gamma=3H$, where $H$ is the Hubble parameter,
but at a later stage of the universe larger $\Gamma$
can be relevant, depending on the
interaction processes considered \cite{relax}. As shown be Sikivie et al. \cite{bec}, axions at the current stage of the universe
are most likely to form a Bose-Einstein
condensate (BEC), which opens up the possibility to exploit macroscopic quantum effects
of the axion condensate
for detection purposes.
The typical parameter ranges that are allowed for QCD dark matter axions are
$
6 \cdot 10^{-6}eV \leq m_ac^2 \leq 2 \cdot 10^{-3} eV
$
and
$
3 \cdot 10^9 GeV \leq f_a \leq 10^{12} GeV
$.
The
product $m_ac^2f_a$ is expected to be of the order $m_ac^2 f_a \sim 6 \cdot 10^{15} (eV)^2$.

Let us consider as a suitable detector a Josephson junction (JJ) \cite{josephson}, which we initially
treat in the approximation of the RSJ model \cite{tinkham} (later we will come to
more specificic physics for S/N/S junctions).
In the RSJ model the phase difference $\delta$
of a JJ driven by a bias current $I$ satisfies
\begin{equation}
\ddot \delta +\frac{1}{RC} \dot{\delta} +\frac{2eI_c}{\hbar C} \sin \delta = \frac{2e}{\hbar C} I \label{2}
\end{equation}
where $I_c$ is the critical current of the junction, and $R$ and $C$ are the normal resistance
and capacity of the junction, respectively.

As pointed out in \cite{beck1,beck2}, the equations of motions for axions (\ref{1}) and
JJs (\ref{2}) are formally identical, and also the numerical values of the coefficients
in the equations are of similar order of magnitude (see \cite{beck1} for some numerical examples).
In this formal analogy the axion mass parameter squared essentially corresponds to the critical current $I_c$,
the product $\vec{E} \cdot \vec{B}$ corresponds to the bias current $I$, and the damping $\Gamma$
corresponds to $(RC)^{-1}$.
%\begin{eqnarray}
%\Gamma &=& \frac{1}{RC} \label{3} \\
%\frac{m_a^2c^4}{\hbar} &=& \frac{2eI_c}{C} \label{4} \\
%\frac{g_\gamma}{\pi f_a^2} c^3 e^2 \vec{E} \vec{B} &=& \frac{2e}{\hbar C} I \label{5}
%\end{eqnarray}
%The
%order of magnitude of typical dark matter axion masses
%numerically also coincides with typical
%values of the quantity $\sqrt{\frac{\hbar 2eI_c}{C}}$
%in Josephson junctions (see \cite{beck1,beck2} for numerical examples).
%This suggests the possibility of an interaction between the phase variables $\theta$ and $\delta$
%under suitable conditions. Such an interaction does not disturb the dark matter status
%of 'invisible' QCD dark matter axions, since it would occur in Josephson junctions only and almost all
%of the ordinary matter contents of the universe is {\em not} in the form of Josephson junctions.

Now let us consider an axion with misalignment angle $\theta$ that enters the weak link region
of a JJ with phase difference $\delta$ (Fig.~1a).
%For convenience we choose the
%initial time point such that initially $\delta =0$.
\begin{figure}
\epsfig{width=8cm,height=5cm,file=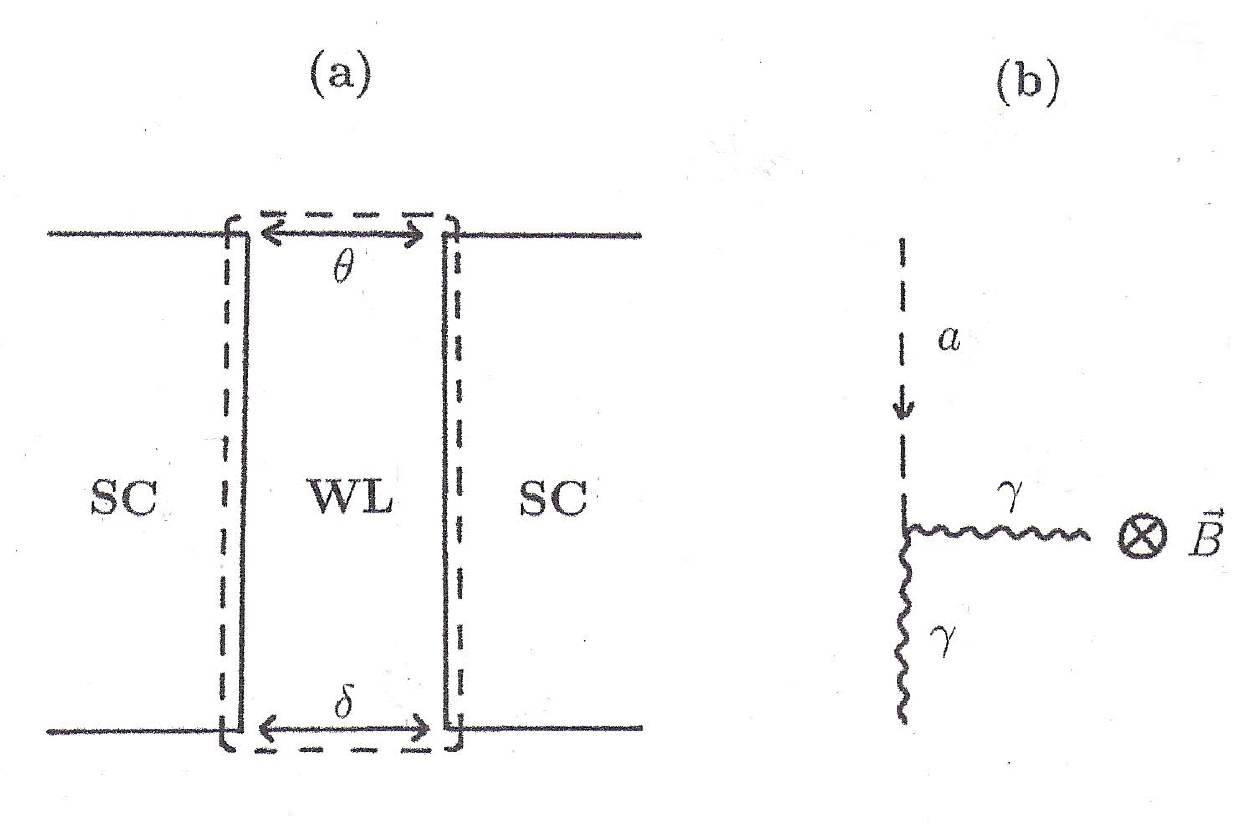}
\caption{(a) Closed integration curve (dashed line) over axion and Josephson phase angles in the weak-link region (WL) of a JJ.
(b) Feynman graph underlying axion-photon decay in a JJ.}
\end{figure}
Both the axion (as a BEC with an equation of motion
identical to a JJ) and the Josephson junction (as a coherent state
of two superconductors (SC) separated by a weak link (WL)) are macroscopic quantum systems
and can be described by a joint macroscopic wave function $\Psi$
in the vicinity of WL.
Similar
to the case where two JJs are put together in a SQUID configuration \cite{tinkham}
the phase variable $\varphi$ of this
wave function $\Psi=|\Psi|e^{i\varphi}$ must be single-valued.
This means that for a given closed integration curve covering the interior of both superconductors (SC)
and the weak link region (WL) one has
%including the interior region of
%the two superconductors (SC), the weak link region of the
%Josephson junction with phase difference $\delta$  and the axion
%with phase difference $\theta$, one has
\begin{equation}
\int_{SC} \nabla \varphi \cdot d \vec{s} + \delta + \theta = 0 \mod 2\pi \label{loop}
\end{equation}
(see Fig.~1a). Here we assume that $\theta$ (thought to be of order $10^{-19}$) enters the
loop as a tiny phase difference over WL, just as a 2nd JJ does.
Condition (\ref{loop}) implies that
$\delta$ and
$\theta$ are no longer independent of each other but influence each other.
Our physical interpretation is
that the incoming axion produces a small perturbation $\theta$ in the CP symmetry status of the two
SCs separated by WL, to which the JJ reacts by building up a small response phase $\delta$
so that CP symmetry is restored.

In the presence of a vector potential $\vec{A}$ we may define gauge-invariant phase differences
$\gamma_i$ ($i=1,2$) by
\begin{eqnarray}
\gamma_1&:=& \delta - \frac{2\pi}{\Phi_0} \int_{weak\; link\; 1} \vec{A} \cdot d \vec{s},  \label{a} \\
\gamma_2&:=& \theta - \frac{2\pi}{\Phi_0} \int_{weak\; link\; 2} \vec{A} \cdot d \vec{s}.  \label{b}
\end{eqnarray}
Here
$\Phi_0=\frac{h}{2e}$ denotes the flux quantum and
{\em weak link 1} denotes the link from the right to the left SC at the bottom of Fig.~1a, whereas
{\em weak link 2} denotes the link from the left to the right SC at the top of Fig.~1a.
For practical purposes it is more convenient to define all links in the same direction (say from the left SC to the
right SC),
% at the position of the axion, respectively the responding Cooper pair.
and the standard formalism employing uniqueness of the phase $\varphi$ modulo $2\pi$
(see, e.g., \cite{tinkham}) then yields the relation
\begin{equation}
\gamma_1 -\gamma_2=2\pi \frac{\Phi}{\Phi_0} \mod 2\pi, \label{flux}
\end{equation}
where $\Phi$ is the magnetic flux through the area enclosed by the chosen closed line of integration.
Eq.~(\ref{flux}) physically means that $\gamma_1$ and $\gamma_2$ synchronize.

Now consider a JJ in the voltage stage, where
%a voltage $V$ is maintained between the
%two superconducting electrodes of the junction. In this case
%the relation between bias current $I$ and applied voltage $V$ is
%\begin{equation}
%V= R \sqrt{I^2-I_c^2} \approx RI \mbox{$\;\;\;$ for $I>>I_c$}.
%\end{equation}
%and the first and third term in eq.~(\ref{2}) can be neglected, so that in good approximation
%\begin{equation}
%\dot{\delta}=\frac{2eRI}{\hbar} = \frac{2eV}{\hbar} \label{xxx}
%\end{equation}
%which means $\delta$ grows linearly in time,
$
\delta (t) = \delta
 (0) +\frac{2eV}{\hbar} t $.
There is an oscillating supercurrent $I_c\sin \delta(t)$ and the junction emits Josephson radiation with frequency
$\hbar \omega_J =2eV$.
%Uniqueness of the joint axion-Josephson wave function implies via
Eqs.~(\ref{a}), (\ref{b}), (\ref{flux}) imply that
$\theta(t)=\delta (t)+const$
(where the constant depends on the magnetic flux included in the loop), meaning that (classically, in WL) the axion
misalignment angle evolves in the same way
as the Josephson phase difference, up to a constant. In particular,
if $\delta$ increases linearly in time, then also $\theta$ is forced to increase linearly in time
with the same rate as $\delta$ does. We thus get
\begin{equation}
\dot{\theta} =\dot{\delta} =\frac{2eV}{\hbar} \label{10}
\end{equation}
from the phase synchronisation condition (\ref{flux}) and
\begin{equation}
\dot{\theta} = -\frac{g_\gamma}{4\pi^2}
\frac{1}{\Gamma f_a^2} c^3 e^2 \vec{E} \vec{B} \label{11}
\end{equation}
from the original equation of motion (\ref{1}) of the axion, neglecting the first and third
term. The joint validity of Eq.~(\ref{10}) and (\ref{11}) implies that the
JJ environment effectively simulates to the axion the existence of a large
non-zero product $\vec{E} \cdot \vec{B}$ in the weak-link region. Using
%The interpretation is that according to its own equation of motion,
%the same effect on the movement of the axion
%misalignement angle could be achieved if there were a huge product $\vec{E}\cdot \vec{B}$
%acting on the axion as given by eq.~(\ref{11}).
$|\vec{E}|=\frac{V}{d}$,
where $d$ is the distance between the superconducting
electrodes of the JJ, we get from (\ref{10}) and (\ref{11})
a {\em formal} magnetic field given by
\begin{equation}
B = \frac{8 \pi^2 \Gamma f_a^2 d}{g_\gamma \hbar c^3 e}. \label{B-field}
\end{equation}
Note that the result (\ref{B-field}) is independent of the applied voltage $V$. The direction of this
effective $\vec{B}$-field is in the same direction as that of $\vec{E}$, i.e.\  it is orthogonal
to the superconducting plates the junction.

%For the current value
%of the Hubble constant $H=2.36 \cdot 10^{-18} s^{-1}$ and a typical size
%of a Josephson junction tunneling region of order $d=2.5 \cdot 10^{-6}m$, one obtains $\vec{B}$-fields
%of the order $10^6 \;T$, the precise value of course depending on $f_a$.
%Such a large $\vec{B}$ field is not producable in a lab in a static way.
%Alternatively, if we substitute for $H$ the value $1/(3RC)$
Putting in typical values for the QCD axion coupling $f_a$ and the
distance $d$, one gets huge
numerical values for $B$, many orders of magnitude higher than what can be achieved
by externally producing a $B$ field in the lab. As a numerical example,
for a typical tunnel junction $d=10^{-9}m$ and $RC \sim 10^{-12}s$ \cite{koch}. Assuming
$f_a \sim 5 \cdot 10^{19}eV=8 J$
and $\Gamma \sim (RC)^{-1}$ one gets $B\sim 10^{34} T$, an incredibly large value.
Much smaller choices of $\Gamma$, of the order of the current Hubble parameter $H\sim 10^{-18}s^{-1}$, still yield
a big $B\sim 10^3T$.

%Our conclusion is that an axion that phase-synchronizes in a Josephson
%junction moves as if there is
%an extremely large $\vec{B}$-field.
%This will make the axion decay immediately.
%Thus phase-synchronised axions cannot exist in the weak link region of a Josephson
%junction, they are expected to decay immediately. This opens up new possibilities
%to develop axion detectors.

Our conclusion is that a phase-synchronised axion cannot exist in the junction but decays
into microwave photons (Fig.~1b).
To roughly estimate the probability $P_{a\to \gamma}$ of this to happen, we may use the
well-known formula from the Primakoff effect \cite{reso}
\begin{equation}
P_{a\to \gamma} = \frac{1}{4\beta_a}(g\; Bec\; L)^2 
\frac{1}{4 \pi \alpha} \left( \frac{sin \frac{qL}{2\hbar}}{\frac{qL}{2\hbar}} \right)^2
\end{equation}
where $q$ is the axion-photon momentum transfer, $\beta_a=v/c$ the axion velocity, $L$ the length of the detector,
and $g:= \frac{g_\gamma \alpha}{\pi f_a}$, where $\alpha$ is the fine structure constant. Inserting the formal value
of the $B$-field (\ref{B-field}) one obtains for $qL << 2 \hbar$
\begin{equation}
P_{a\to \gamma} = \frac{1}{\beta_a \hbar^2 c^4} (f_a \Gamma d L)^2 4 \pi \alpha
\end{equation}
In particular, $P_{a \to \gamma}=1$ corresponds to a very short length scale, namely
\begin{equation}
L= \frac{\hbar c^2}{\sqrt{4 \pi \alpha}} \sqrt{\beta_a} \frac{1}{f_a \Gamma d}
\end{equation}
For our previous numerical example and an axion velocity of $v=2.3 \cdot 10^5 m/s$ one gets $L\sim 10^{-22}m$,
which means that
the axion immediately decays at the surface of the weak-link region.
Moreover, since $P_{a\to \gamma}=P_{\gamma\to a}$ it can
equally likely recombine back into an axion when leaving the weak-link region, thus producing ATJ.
% An axion seeing
%a magnetic field of size given by (\ref{B-field}) decays with probability 1
%while passing the weak region of the junction, induced by Josephson radiation.
%The
%direction of this virtual $B$-field simulated by the biased JJ
%is collinear to the $\vec{E}$-field in the junction. Its size is larger by a factor $\sim 10^5$ compared
%to typical magnetic field strengths that can be achieved in laboratory experiments.
%Upon synchronization,
%the JJ environment thus simulates the presence of a very large (virtual) magnetic field to the
%axion, which is collinear to the electric field and which makes the axion decay immediately.
%
%We have thus shown that JJs act like a `wall' for phase-synchronized axions, and axions are not expected to exist within
%the weak-link region of a JJ but decay into microwave photons.
The total probability of axion decay
in the junction is given by $pP_{a\to \gamma}$, where $p$ is the probability that the axion
phase synchronizes. There are no astrophysical or cosmological constraints on $p$ since almost all
matter of the universe is not in the form of JJs, hence $p\sim O(1)$ is compatible with the
astrophysical dark matter status of the axion.

Let us now come to measurable effects to test this theoretical idea.
It is well-known that external
microwave radiation applied to a Josephson junction leads to distortions in the $I-V$ curve,
the well-known Shapiro steps \cite{shapiro}. Within the RSJ model, one can calculate
(see, e.g., \cite{chen})
that external monochromatic microwave radiation of signal frequency $\hbar \omega_s=2eV_s$ leads to a distortion $I_s$ in the
measured current-voltage curve $I(V)$ given by
\begin{equation}
\hspace{-7.5cm} I_s(V)= \nonumber
\end{equation}
\vspace{-0.5cm}
\begin{equation}
\frac{P_s}{4} (RI_c)^2 \frac{1}{V^2} \left[ \frac{V+V_s}{(V+V_s)^2 +(\frac{\delta V}{2})^2}
+ \frac{V -V_s}{(V-V_s)^2 +(\frac{\delta V}{2})^2} \right] .
\end{equation}
Here
$P_s$ is the signal power, and $\delta V$ is determined by the line width of the Josephson frequency.

When the Josephson frequency $\hbar \omega_J=2eV$ resonates with the axion mass $m_ac^2$
we may assume that all axions hitting the weak-link region synchronize and decay ($p=1$), hence
the expected signal power $P_s$ is given by
\begin{equation}
P_s=\rho_{a} v A.
\end{equation}
Here $\rho_a$ is the Halo axionic dark matter energy density close to the earth, $v$ is the velocity
of the earth relative to the galactic center, and $A$ is the area of the weak-link region perpendicular to
the axion flow. Neglecting fluctuations, the velocity $v$ is known to be approximately $2.3 \cdot 10^5 m/s$,
with a yearly modulation of about 10\% around its mean value \cite{lab1,bernabei}.

%For a typical tunnel junction \cite{josephson,koch}, the distance between the
%superconducting electrodes is $d \sim 1nm$ and hence $A\sim 1nm \times 1\mu m \sim 10^{-15}m^2$,
%whereas S/N/S junctions \cite{hoffmann,lhotel,hoss,dubos}
%allow for a much larger area of the weak link region, $A \sim 1\mu m \times 1 \mu m \sim 10^{-12} m^2$,
%since for these junctions $d\sim 1 \mu m$ is much larger.
%Thus we expect for S/N/S junctions an axion signal that is stronger by a factor 1000, and which in addition
%can be further amplified by multiple Andreev reflections.
%We are thus pointed towards S/N/S junctions as being particularly suited for
%axion detection.

%However, for these junctions the RSJ model is only a crude
%approximation.
%The weak link region
%of S/N/S junctions consists of a normal
%metal plate and the corresponding area is much larger
%than for tunnel junctions. Cooper pairs do no tunnel
%the weak-link region but their transport through the weak-link region is mediated by multiple Andreev reflection of hole-electron %pairs
%(\cite{x}).

We propose as a specific
microscopic model underlying the interaction of axions with
resonant S/N/S junctions the process sketched in Fig.~2.
\begin{figure}
\epsfig{width=8cm,height=6cm,file=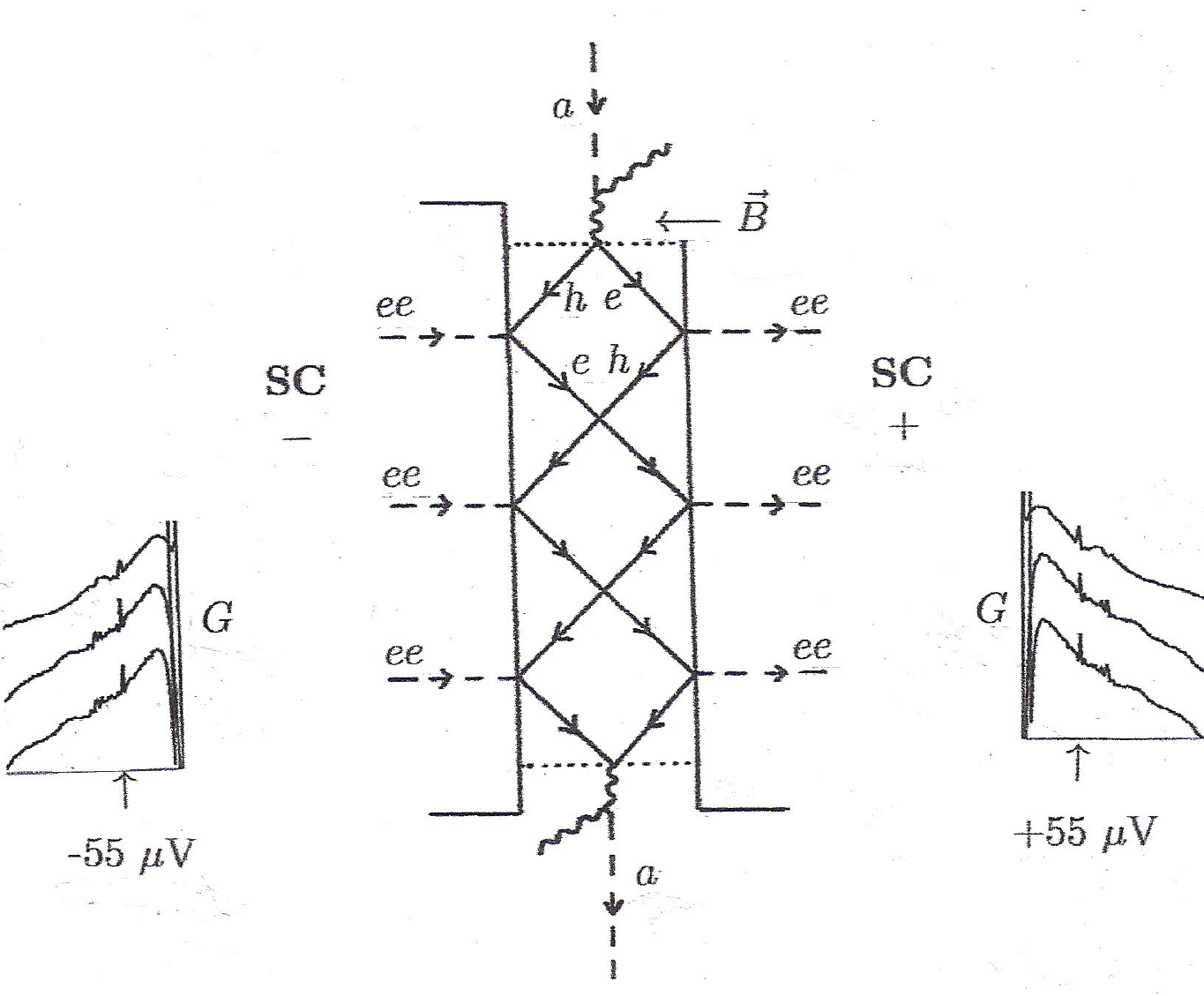}
\caption{Axion triggering the transport of $n$ Cooper pairs $ee$ in an S/N/S junction by multiple
Andreev reflections, here shown for the example $n=3$. The dotted line corresponds
to the normal metal surface. The left and right insets show the
shape of the differential conductance curve
$G(V)$ as measured by Hoffmann et al. \cite{hoffmann} for $T=0.9K,0.5K,0.1K$ (top to bottom), with a peak occuring
at $\pm 0.055mV$.}
\end{figure}
An axion entering the weak link region with a
transversal velocity component
decays close to the normal metal surface
into two microwave photons, one with $q\approx 0$
and the other one with frequency $\hbar \omega_s \approx m_ac^2=2eV_s$. The
$q \approx 0$ photon transfers its small momentum to a hole-electron pair in the weak-link
region (holes $h$ and electrons $e$
can enter from the surrounding superconductors with energy $-\Delta$, where $\Delta$ is
the gap energy).
The hole and the electron perform multiple Andreev
reflections in the usual way \cite{hoffmann,lhotel}, meaning the hole is reflected at the S/N interface
as an electron and annihilates in this process a Cooper pair, whereas the
electron is reflected at the other S/N interface as a hole, creating in this
process a Cooper pair. In total, there are $n$ Andreev reflections, with
\begin{equation}
n \approx \frac{2\Delta}{eV}+1
\end{equation}
\cite{hoffmann}. At the end of this process,
when both the electron and hole energy exceed the gap energy $\Delta$
they either just leave the weak-link region or
annihilate back into a low-energy photon, which together with another
photon of Josephson frequency $\hbar \omega_J=2eV_s=m_ac^2$ can recombine
back into an axion, which leaves the detector unharmed. In total, there
is an ATJ process and each
incident axion triggers the transport of $n$ Cooper pairs.
These additional Cooper pairs produce a signal
$G_s=dI_s/dV$
in the measured differential conductance $G(V)$ of the junction
at the signal voltage $V_s=m_ac^2/(2e)$. The total signal current
produced by axions is given by
\begin{equation}
I_s= \int G_s dV= \frac{N_a}{\tau} \cdot n \cdot 2e = \frac{\rho_a}{m_ac^2} v A  \cdot n \cdot 2e \label{axion-density}
\end{equation}
where $N_a/\tau$ is the number of axions hitting the normal metal region per time unit $\tau$.
Since $2eV_s=m_ac^2$ we get
\begin{equation}
\rho_a=\frac{I_s V_s}{vAn}. \label{amman}
\end{equation}
This can be used to experimentally estimate the axion dark matter density $\rho_a$ from
an experimental measurement of $V_s$ and $I_s$.
%The line width of the conductivity peak is expected to be of order $1/(2n)$
%due to the uncertainty in the integer number $n$ of Andreev reflections
%taking place, which produces a voltage uncertainty of the
%order $RI_s\cdot \frac{1}{2n}$.

In \cite{hoffmann} Hoffmann et al. have observed a signal
of unknown origin that is consistent with our theoretical expectations. Independent
of the temperature (which is varied from 0.1K to 0.9K) they consistently
observe a small peak in their measured differential conductance $G(V)$ at the
voltage $V_s=\pm 0.055mV$ (see insets of Fig.~2, data from \cite{hoffmann}). Their measurements
provide evidence for a signal
current feature of size $I_s=(8.1\pm 1.0) \cdot 10^{-8}A$, which is
obtained by integrating the area under the observed signal peak of the differential
conductance. Their noise measurements also indicate
that every quasi-particle performs $n=7$ Andreev reflections \cite{hoffmann}.
The area of the metal plate of their junction
is $A=0.85 \mu m \times 0.4 \mu m= 3.4 \cdot
10^{-13}m^2$. From $2eV_s=m_ac^2$ we thus obtain an axion mass
prediction of $m_ac^2=110 \mu eV$ (equivalent to $f_a\sim 5.5 \cdot 10^{10}GeV$), and
eq.~(\ref{amman}) yields the prediction
$\rho_a= (0.051\pm 0.006) GeV/cm^3$.

Astrophysical observations suggest that the galactic dark matter density $\rho_d$
near the earth is about $\rho_d= (0.3\pm 0.1) GeV/cm^3$ \cite{dm-density}. But this includes all kinds of
dark matter particles,
including WIMPS. Generally, axions of high mass will make up only a fraction
of the total dark matter density of the universe, which
can be estimated from cosmological considerations to be about
$\rho_a/\rho_d \approx (24 \mu eV/ m_ac^2)^{7/6}$ \cite{duffy}.
%Whereas axions of mass of about $24 \mu eV$ could make up
%for the entire dark matter contents
%of the universe,
For
$m_ac^2 =110 \mu eV$
we thus expect an axionic dark matter density that is
a fraction
$(24/110)^{7/6}\approx 0.17$ of the total dark matter density, giving
$\rho_a\approx 0.17  \cdot \rho_d
= (0.051 \pm 0.017) GeV/cm^3$. The experimental results of Hoffmann et al.
together with our theoretical prediction (\ref{amman})
are thus in perfect agreement with what is expected from astrophysical observations.

To either refute or confirm our hypothesis that the signal seen in \cite{hoffmann}
is produced by dark matter axions, further measurements are needed. Clearly
one should test if the signal
survives careful shielding of the junction
from any external microwave radiation. A signal produced by axions cannot
be shielded. Moreover, one might look for a possible dependence
of the measured signal intensity on the spatial orientation of the metal
plate relative to the galactic axion flow (a precise directional measurement
would be extremely helpful).
Finally, and
most importantly, the velocity $v$ by which the earth moves through the
axionic BEC of the galactic halo exhibits a yearly modulation of about
10\%, with a maximum in June and a minimum in December
(as used in 
searches for WIMPS \cite{bernabei}). If the JJ
signal is produced by axions, then its intensity should show the same
10\% modulation effect over a period of a year.
This can be tested in future experiments. In \cite{hoffmann}
also some fine structure of $G(V)$ near 0.08mV is observed,
which might be a hint for further axion-like particles
with different mass.

To conclude, in this letter we have described a macroscopic quantum effect
in Josephson junctions that may help
to prove the existence of axionic dark matter in future measurements.
Phase-synchronised
axions cannot exist in the weak-link region of JJs
due to a (formal) huge magnetic field that is simulated to them by the
driven JJ environment in the voltage stage.
Axions are expected to decay when entering the
weak-link region of the junction and trigger
the transport of additional Cooper pairs.
This leads to a small measurable signal
for the differential conductance
(a Shapiro step-like signal without externally applied microwave radiation)
if the axion mass
resonates with the Josephson frequency. The effect is particularly
strong in S/N/S junctions which have a much larger weak-link region
than tunnel junctions and where the Cooper pair transport is amplified
by multiple Andreev reflections. A candidate signal of
unknown origin has been observed in
measurements of Hoffmann et al.
\cite{hoffmann}, which interpreted in this way points to an axion
mass of 0.11 meV and a local axionic energy density of 0.05 GeV/$cm^3$.
%Further
%experimental measurements are needed to either confirm or
%refute the hypothesis that this signal is produced by axions.
%Our calculations show that the intensity of the observed signal is consistent with the expected
%dark matter axion flow rate through the junction.

\end{document}